\documentclass[aps,pre,twocolumn,superscriptaddress]{revtex4-1}

\usepackage[pdftex]{graphicx}
\usepackage{bm}
\usepackage{amsmath}
\usepackage{color}

\begin{document}


\title{Diffusion with a broad class of stochastic diffusion coefficients}

\author{Go Uchida}
\affiliation{Department of Mechanical Systems Engineering, 
Tokyo Metropolitan University, Tokyo 1920397, Japan}
\affiliation{Graduate School of Information Science, 
University of Hyogo, Hyogo 6500047, Japan}
\author{Hitoshi Washizu}
\affiliation{Graduate School of Information Science, 
University of Hyogo, Hyogo 6500047, Japan}
\author{Hiromi Miyoshi}
\affiliation{Department of Mechanical Systems Engineering, 
Tokyo Metropolitan University, Tokyo 1920397, Japan}


\date{\today}

\begin{abstract}
In many physical or biological systems, 
diffusion can be described by Brownian motions with 
stochastic diffusion coefficients (DCs). 
In the present study, 
we investigate properties of the diffusion with a broad class of stochastic DCs 
with a novel approach. 
We show that 
for a finite time, 
the propagator is non-Gaussian and 
heavy-tailed. 
This means that when the mean square displacements are the same, 
for a finite time, 
some of the diffusing particles with stochastic DCs diffuse farther than 
the particles with deterministic DCs or exhibiting a fractional Brownian motion. 
We also show that when a stochastic DC is ergodic, 
the propagator converges to a Gaussian distribution in the long time limit. 
The speed of convergence is determined by the autocovariance function of the DC.  
\end{abstract}


\maketitle

\section{Introduction}
Diffusion 
in physical systems or in biological systems 
is often described by 
Brownian motions with stochastic diffusion coefficients (DCs). 
For example, 
ligands often fluctuate between different conformational states. 
In the models of ligand diffusion, 
different DCs are given 
for the different states~\cite{Reingruber09,Reingruber10,Bressloff17}. 
An experimental observation of 
subdiffusion of a certain receptor on a cell membrane is also explained by a model of a Brownian motion with a stochastic DC~\cite{Manzo15}. 
In this model, 
the stochasticity of the DC comes from dynamic heterogeneity of the cell membrane~\cite{Manzo15}. 

It has been shown that for a specific class of and specific models of Brownian motions with stochastic DCs, 
the propagators are heavy-tailed for short times~\cite{Chechkin17, Metzler20,Miyaguchi16}. 
When a DC is described by the square of an Ornstein-Uhlenbeck process (minimal diffusing-diffusivity model), 
the propagator is a Laplace-like distribution with exponential tails for short times~\cite{Chechkin17, Metzler20}.  
When the DC fluctuates between two states and 
the distributions of sojourn time in each state are given by power-law distributions (two-state model with power-law distributions),   
the propagator is heavy-tailed for short times~\cite{Miyaguchi16}.  
However, 
unlike continuous time random walk (CTRW) where the exponential decay is shown to be a common property of propagators~\cite{Barkai20}, 
it remains unknown whether a heavy-tailed propagator is a common property of Brownian motions with stochastic DCs.  

There is another open question about diffusion with stochastic DCs. 
For specific models of stochastic DCs, 
theoretical studies have shown that at long times, 
heavy-tailed propagators cross over to Gaussian propagators. 
When a DC follows a certain random walk, 
the propagator is exponential at short times and Gaussian at long times~\cite{Chubynsky14}. 
For the minimal diffusing-diffusivity model, 
it has been shown that at times longer than some correlation time of the DC, 
a crossover from a heavy-tailed propagator to a Gaussian propagator occurs~\cite{Chechkin17}.  
However, 
as with a property of propagators for short times, 
it remains unknown how broad a class of stochastic DCs the crossover occurs. 

In the present study, 
we investigate properties of the diffusion 
with a broad class of stochastic DCs, with a particular focus on the propagator.  
In Sec.~\ref{sec:model}, 
we describe 
an overdamped Langevin equation with 
a stochastic DC. 
The diffusion with a stochastic DC can be described by 
an overdamped Langevin equation with a stochastic DC. 
In Sec.~\ref{sec:FPe}, 
we present a novel approach to find an expression for the propagator. 
We derive the diffusion equation corresponding to the overdamped Langevin equation with a stochastic DC and 
then derive an expression for the propagator 
by solving the diffusion equation.   
In Sec.~\ref{sec:propdiff}, 
we reveal properties of the diffusion through properties of the propagator. 
In Sec.~\ref{sec:discuss}, 
we discuss  
the relation of our approach to other approach and compare our results with those of other models.

\section{Model} \label{sec:model}
The overdamped Langevin equation with a stochastic DC 
is given by 
\begin{equation}
    \frac{d\bm{x}(t)}{dt} = \sqrt{2D(t)} \bm{\xi }(t),  \label{eq:pifgt1}
\end{equation}
where 
$ \bm{x}(t) $ is the $n$-dimensional position of the diffusing particle and 
$ \bm{x}(t) = (x_{1}(t),x_{2}(t),\cdots ,x_{n}(t))^{T} $. 
In Eq.~(\ref{eq:pifgt1}), 
$ D(t) $ is a stochastic process and represents a DC. 
In Eq.~(\ref{eq:pifgt1}), 
$ \bm{\xi }(t) $ is a vector of Gaussian white noises and 
$ \bm{\xi }(t) = \left( \xi _{1}(t), \xi _{2}(t), \cdots ,\xi _{n}(t) \right)^{T} $: $ \langle \bm{\xi }(t) \rangle = \bm{0} $ and 
$ \langle \xi _{i}(t) \xi _{j}(t') \rangle = \delta _{ij}\delta(t-t') \, (i,j=1,2,\cdots ,n) $. 
We assume that $ D(t) $ and $ \xi _{i}(t) $ are statistically independent.

\section{Diffusion equation} \label{sec:FPe}
\subsection{Derivation}
Here, 
we derive the diffusion equation corresponding to Eq.~(\ref{eq:pifgt1}). 
The stochastic integral equation corresponding to Eq.~(\ref{eq:pifgt1}) is given by 
\begin{equation}
    \bm{x}(t) = \bm{x} _{0} + \int_{0}^{t} \sqrt{2D(t')} d\bm{B}(t') ,    \label{eq:sie} 
\end{equation}
where $ \bm{x} _{0} = \bm{x}(0) $, $ \bm{B}(t) $ is a vector of Wiener processes, and 
$ \bm{B}(t) = \left( B _{1}(t), B _{2}(t), \cdots ,B _{n}(t) \right)^{T} $: 
$ \langle d\bm{B}(t) \rangle = \bm{0} $ and 
$ \langle dB _{i}(t) dB _{j}(t) \rangle = \delta _{ij}dt $. 

Eq.~(\ref{eq:pifgt1}) is just a formal equation. 
If $ D(t) $ is deterministic and $ \int_{0}^{t} D(t') dt' < \infty $, 
the integral on the right hand side of Eq.~(\ref{eq:sie}) is the Wiener integral. 
If $ D(t) $ is a stochastic process defined on the measurable space on which $ \bm{B}(t) $ is defined, 
the integral on the right hand side of Eq.~(\ref{eq:sie}) can be the stochastic integral.   
However, 
in this study, 
$ D(t) $ is neither. 
Thus, 
for further calculations, 
we need to give an interpretation of Eq.~(\ref{eq:pifgt1}). 

One natural interpretation of Eq.~(\ref{eq:pifgt1}) is given by the equation: 
\begin{eqnarray}
    \bm{x}(t;\omega ) &=& \bm{x} _{0} + \int_{0}^{t} \sqrt{2D(t',\omega )} d\bm{B}(t') \; (\omega \in \Omega),    \label{eq:siei} \\ & &\int _{0}^{t} D(t',\omega ) dt' < \infty, \label{eq:sint} 
\end{eqnarray}
where $ \Omega $ represents the sample space for $ D(t) $, $ \omega $ represents a sample, 
$ D(t,\omega ) $ represents a sample path of $ D(t) $, and 
$ \bm{x}(t;\omega ) $ is the position of the diffusing particle for a given sample path $ D(t,\omega ) $. 
Each sample path of $ D(t) $ is deterministic. 
Thus, 
the integral in the right hand side of Eq.~(\ref{eq:siei}) is the Wiener integral.  
The condition given by Eq.~(\ref{eq:sint}) is natural because $ 0 \le D(t) < \infty $. 

The diffusion equation corresponding to Eq.~(\ref{eq:siei}) is given by 
\begin{equation}
    \left[\partial _{t} -D(t;\omega ) \nabla ^{2} \right] G \left(\bm{x},t,\bm{x}_{0};\omega \right) = \delta (\bm{x}-\bm{x}_{0})\delta (t) \; (\omega \in \Omega), \label{eq:tdfpgt}
\end{equation}
where $ G \left(\bm{x},t,\bm{x}_{0};\omega \right) $ represents the propagator for a given sample path $ D(t;\omega ) $.

\subsection{Propagator} \label{sec:nGGf}
Here, 
we solve Eq.~(\ref{eq:tdfpgt}) and derive the propagator. 
Under the natural boundary condition, 
the solution of Eq.~(\ref{eq:tdfpgt}) is given by~\cite{Molini11} 
\begin{equation}
    G (\bm{x},t,\bm{x}_{0};\omega ) = \frac{1}{\left[4\pi S(t;\omega )t \right]^{n/2}} \exp \left[ -\frac{\left| \bm{x} -\bm{x}_{0} \right|^2}{4S(t;\omega )t} \right] , \label{eq:rhogfgtt}
\end{equation}
where 
$ S(t;\omega ) $ represents the time average of $ D(t;\omega ) $ 
and 
is given by 
\begin{equation}
    S(t;\omega ) = \frac{1}{t} \int _{0}^{t} D(t';\omega ) dt'. \label{eq:Mgt}
\end{equation}

The propagator $ G (\bm{x},t,\bm{x}_{0};\omega ) $ depends on $ \omega $ only through  
$ S(t;\omega ) $. 
Thus, 
the propagator for the diffusion described by Eq.~(\ref{eq:pifgt1}) 
$ G (\bm{x},t,\bm{x}_{0}) $ is 
given by 
\begin{equation}
    G (\bm{x},t,\bm{x}_{0}) = \int _{0}^{\infty }  \frac{p(S,t)}{\left(4\pi St \right)^{n/2}} \exp \left( -\frac{\left| \bm{x} -\bm{x}_{0} \right|^2}{4St} \right) dS, \label{eq:pdexp2gf}
\end{equation}
where $ p(S,t) $ is the probability distribution of the time average of the DC (TADC), $ S(t) $. 

From Eq.~(\ref{eq:pdexp2gf}), 
we can see that the propagator is unimodal and 
the peak is at $ \bm{x} = \bm{x} _{0} $. 
In addition, 
when $ 0 < D(t) $,  
the propagator is differentiable at $ \bm{x} = \bm{x} _{0} $. 
When $ D(t) $ can take the value of zero, 
the propagator can be non-differentiable at $ \bm{x} = \bm{x} _{0} $ if 
$ t $ is smaller than the time that characterizes the change of $ D(t) $ (see Ref.~\cite{Soria21} for an example).   
However, 
even if $ D(t) $ can take the value of zero, 
the propagator is differentiable at $ \bm{x} = \bm{x} _{0} $ if 
$ t $ is larger than the time that characterizes the change of $ D(t) $.   
The differentiability at the peak means that the peak is not sharp. 
The smoothness of the peak is important in comparing with other diffusion models (see Sec.~\ref{sec:discuss}).

\section{Properties of the diffusion} \label{sec:propdiff}
\subsection{General theory}
Here, 
we reveal properties of the diffusion with a broad class of stochastic DCs 
through properties of the propagator. 
From Eq.~(\ref{eq:pdexp2gf}), 
we can derive expressions for the moments of all orders. 
From Eq.~(\ref{eq:pdexp2gf}), 
we have 
\begin{equation}
    \int _{\bm{x} } \exp \left( \lambda \left| \delta \bm{x} \right|^2 \right)G(\bm{x},t,\bm{x}_{0}) d\bm{x} = \int _{0}^{\infty } \frac{p(S,t)}{(1-4St\lambda )^{n/2}} dS, \label{eq:mgfxmgfS}
\end{equation}
where $ \delta \bm{x} $ represents the displacement and is given by $ \delta \bm{x} = \bm{x} - \bm{x}_{0} $, 
and $ \int _{\bm{x} } d\bm{x} = \int _{-\infty }^{\infty } \cdots \int _{-\infty }^{\infty } dx_{1}\cdots dx_{n} $.  
For the left-hand side of Eq.~(\ref{eq:mgfxmgfS}), 
substituting $ \lambda = 0 $ into 
the $k$-th order derivative of $ \lambda $ leads to 
\begin{equation}
    \left. \frac{d^{k}}{d\lambda ^{k}}\int _{\bm{x} } \exp \left( \lambda \left| \delta \bm{x} \right|^2 \right)G(\bm{x},t,\bm{x}_{0}) d\bm{x} \right|_{\lambda =0} = \left\langle |\delta \bm{x}(t)|^{2k} \right\rangle , \label{eq:mgfGm}
\end{equation}
where $ k $ is a positive integer. 
In this study, 
the ensemble average $ \langle \, \rangle $ is taken over the ensemble appropriate for the stochastic variable to be averaged. 
For example, 
in Eq.~(\ref{eq:mgfGm}), 
$ \left\langle |\delta \bm{x}(t)|^{2k} \right\rangle $ is given by 
\begin{equation}
    \left\langle |\delta \bm{x}(t)|^{2k} \right\rangle = \int_{\bm{x}} |\delta \bm{x}|^{2k} G(\bm{x},t,\bm{x}_{0}) d\bm{x} . 
\end{equation}   
For the right-hand side of Eq.~(\ref{eq:mgfxmgfS}), 
substituting $ \lambda = 0 $ into 
the $k$-th order derivative of $ \lambda $ leads to 
\begin{eqnarray}
    \left. \frac{d^{k}}{d\lambda ^{k}}\int _{0}^{\infty } \frac{p(S,t)}{(1-4St\lambda )^{n/2}} dS \right|_{\lambda =0} &=& (2t)^{k}n(n+2)\cdots \nonumber \\& &\times (n+2k-2)  \left\langle S^{k} (t) \right\rangle . \nonumber \\ \label{eq:kdS}
\end{eqnarray}
From Eqs.~(\ref{eq:mgfxmgfS}), (\ref{eq:mgfGm}), and (\ref{eq:kdS}), 
we obtain  
\begin{equation}
    \left\langle |\delta \bm{x}(t)|^{2k} \right\rangle = (2t)^{k}n(n+2)\cdots (n+2k-2) \left\langle S^{k} (t) \right\rangle. \label{eq:mSG}
\end{equation}
In addition, 
from Eq.~(\ref{eq:pdexp2gf}), 
we have 
\begin{equation}
    \left\langle |\delta \bm{x}(t)|^{2k-1} \right\rangle = 0. \label{eq:mSG2}
\end{equation}

From Eq.~(\ref{eq:mSG}), 
we have 
\begin{equation}
    \langle |\delta \bm{x}(t)|^{2} \rangle = 2nt\left\langle S (t) \right\rangle . \label{eq:msdms}
\end{equation}
This equation has already been derived for a broad class of stochastic DCs 
using the Langevin formalism [Eq. (24) in Ref.~\cite{Miyaguchi16}]. 

From Eq.~(\ref{eq:msdms}), 
we can see that the mean square displacement (MSD) $ \langle |\delta \bm{x}(t)|^{2} \rangle $ does not reflect fluctuations in DCs. 
From Eq.~(\ref{eq:pdexp2gf}), 
we can see that for a finite time, 
the propagator is non-Gaussian if a DC is stochastic.  
This suggests that the non-Gaussian parameter 
reflects 
fluctuations in a DC. 
The non-Gaussian parameter~\cite{Miyaguchi16,Onuki98,Hofling13} $ A(t) $ 
is given by 
\begin{equation}
    A(t) = \frac{\left\langle \left[ \delta \bm{x} ^{T}(t) \Sigma ^{-1} (t)\delta \bm{x}(t) \right] ^{2} \right\rangle }{n(n+2)} - 1, \label{eq:rAk}
\end{equation}
where $ \Sigma (t) $ represents the variance-covariance matrix. 
When the propagator is Gaussian, 
$ A(t) = 0 $. 
From Eq.~(\ref{eq:mSG}), 
we have 
\begin{equation}
    \Sigma (t) = 2t\left\langle S(t) \right\rangle  I_{n}, \label{eq:S1}
\end{equation}
where $ I_{n} $ is the $n$-th order identity matrix. 
From Eqs.~(\ref{eq:msdms}) and (\ref{eq:S1}), 
we obtain 
\begin{equation}
    \Sigma (t) = \frac{\langle |\delta \bm{x}(t)|^{2} \rangle}{n} I_{n}. 
\end{equation}
Thus, 
we have 
\begin{equation}
    A(t) = \frac{n\langle |\delta \bm{x}(t)|^{4} \rangle}{(n+2)\langle |\delta \bm{x}(t)|^{2} \rangle ^{2}} - 1. \label{eq:kuoimx}
\end{equation}
On the other hand, 
from Eq.~(\ref{eq:mSG}), 
we can obtain the variance of the TADC: 
\begin{eqnarray}
    \left\langle \left( S\left( t \right) - \left\langle S \left( t \right) \right\rangle  \right) ^{2}\right\rangle &=& \left\langle S^{2} \left( t\right) \right\rangle - \left\langle S \left( t \right) \right\rangle ^{2} \nonumber \\ &=& \frac{\langle |\delta \bm{x}(t)|^{4} \rangle}{4t^{2}n(n+2)} - \left\langle S \left( t \right) \right\rangle ^{2}. \nonumber \\ \label{eq:varS}
\end{eqnarray}
From Eqs.~(\ref{eq:msdms}) and (\ref{eq:varS}), 
we have  
\begin{eqnarray}
    \frac{\left\langle \left( S\left( t \right) - \left\langle S \left( t \right) \right\rangle  \right) ^{2}\right\rangle }{\left\langle S \left( t \right) \right\rangle ^{2}} &=& \frac{\langle |\delta \bm{x}(t)|^{4} \rangle}{4t^{2}n(n+2)\left\langle S \left( t \right) \right\rangle ^{2}} - 1 \nonumber \\ &=& \frac{n\langle |\delta \bm{x}(t)|^{4} \rangle}{(n+2)\langle |\delta \bm{x}(t)|^{2} \rangle ^{2}} - 1. \nonumber \\ \label{eq:varS2}
\end{eqnarray}
Thus, 
from Eqs.~(\ref{eq:kuoimx}) and (\ref{eq:varS2}),  
we obtain 
\begin{equation}
    A(t) = \frac{\left\langle \left( S\left( t \right) - \left\langle S \left( t \right) \right\rangle \right) ^{2}\right\rangle }{\left\langle S \left( t \right) \right\rangle ^{2}}. \label{eq:betavarS}
\end{equation}
From this equation, 
we can see that the non-Gaussian parameter is equal to 
the square of the coefficient of variation     
of the TADC. 

When $ t $ is sufficiently smaller than the time that characterizes the change in $ D(t) $, 
from Eq.~(\ref{eq:Mgt}), we have an approximation:
\begin{equation}
    S(\omega ) \approx D(0;\omega ) . \label{eq:Mgt30}
\end{equation} 
Thus, 
for sufficiently small $ t $, 
we have
\begin{equation}
    A(t) \approx \frac{\left\langle \left( D\left( 0 \right) - \left\langle D \left( 0 \right) \right\rangle \right) ^{2}\right\rangle }{\left\langle D \left( 0 \right) \right\rangle ^{2}}.  \label{eq:stAg}
\end{equation}

Here, 
using Eq.~(\ref{eq:betavarS}), 
we show that for a finite time, 
the propagator is heavy-tailed relative to a Gaussian distribution 
when a DC is stochastic. 
From Eq.~(\ref{eq:betavarS}), 
we can see that for a finite time, 
the non-Gaussian parameter is positive when a DC is stochastic. 
The non-Gaussian parameter is essentially equivalent to 
the kurtosis of the propagator. 
The kurtosis is a measure of whether a distribution is heavy-tailed or light-tailed relative to a Gaussian distribution.  
The kurtosis $ \beta (t) $ and the non-Gaussian parameter have the relation (for the kurtosis see~\cite{Mardia70}): 
\begin{equation}
    \beta (t) = n(n+2)\left( A\left( t\right) + 1 \right).    \label{eq:bA} 
\end{equation}
From this equation, 
for a finite time, 
we have 
\begin{equation}
    \beta (t) > n(n+2), \label{eq:pkrts}
\end{equation} 
because $ A(t) > 0 $ for a finite time. 
The kurtosis is equal to $ n(n+2) $ for a Gaussian distribution. 
Thus, 
from Eq.~(\ref{eq:pkrts}), 
we can see that for a finite time, 
the propagator is heavy-tailed relative to a Gaussian distribution. 
This means that when the MSDs are the same, 
some of the diffusing particles with stochastic DCs diffuse farther than 
the particles with deterministic DCs for a finite time.   
Note that from Eq.~(\ref{eq:pdexp2gf}), 
the propagators are Gaussian for deterministic DCs.  
 
From Eq.~(\ref{eq:rhogfgtt}), 
we have 
\begin{equation}
    \langle |\delta \bm{x}(t;\omega )|^{2} \rangle = 2nt S (t;\omega ). \label{eq:msdmssp}
\end{equation}
From this equation, 
we can see that 
the MSD is a stochastic variable. 
From Eqs.~(\ref{eq:betavarS}) and (\ref{eq:bA}), 
we can also see that 
even if the values of 
$ \left\langle S \left( t \right) \right\rangle $ are equal, 
the stochastic process $ D(t) $ with the larger variance of the TADC gives the larger value of the kurtosis.  
This is natural because 
a large variance in a TADC leads to a large variance in an MSD and 
thus some particles diffuse farther.  

Variation in a TADC comes from fluctuations in a DC. 
What properties of a DC lead to larger TADC variance? 
From Eq.~(\ref{eq:Mgt}), 
we have 
\begin{equation}
    \left\langle \left( S\left( t \right) - \left\langle S \left( t \right) \right\rangle  \right) ^{2}\right\rangle = \frac{1}{t^{2}} \int_{0}^{t} \int_{0}^{t} C(t',t'') dt'dt'',    \label{eq:SvarDcorr}
\end{equation}
where $ C(t',t'') $ represents the autocovariance function of $ D(t) $ and is given by 
\begin{equation}
    C(t',t'') = \left\langle \delta D(t') \delta D(t'') \right\rangle . \label{eq:krtg}
\end{equation}
In this equation, 
$ \delta D(t) $ is given by 
$ \delta D(t) = D(t) - \left\langle D(t) \right\rangle  $. 
From Eq.~(\ref{eq:SvarDcorr}), 
we can see that 
the larger the variance of $ D(t) $ and the longer the correlation lasts, 
the larger the variance of a TADC.

\subsection{Normal diffusion}    \label{ss:nd}
When $ \left\langle D(t) \right\rangle $ is a constant,  
from Eq.~(\ref{eq:msdms}), 
we have 
\begin{equation}
    \langle |\delta \bm{x}(t)|^{2} \rangle = 2n\mu t, \label{eq:x2mu}
\end{equation}
where $ \mu = \left\langle D(t) \right\rangle $.  
The diffusion is normal. 

When $ D(t) $ is a stationary process,  
$ \left\langle D(t) \right\rangle $ is a constant,  
Thus, 
when $ D(t) $ is a stationary process, 
the diffusion is normal. 
When $ D(t) $ is a stationary process, 
it can be seen more clearly that the variance of the DC plays an important role in determining 
the magnitude of the fluctuation of TADC and, in turn, the value of $ A(t) $. 
When $ D(t) $ is a stationary process, 
we can find the upper limit of $ A(t) $. 
From Eq.~(\ref{eq:betavarS}), 
when $ D(t) $ is a stationary process, 
we have
\begin{equation}
    A(t) = \frac{\left\langle \left( S\left( t \right) - \mu \right) ^{2}\right\rangle }{\mu ^{2}}. \label{eq:betadvarS}
\end{equation}
In addition, 
we have 
\begin{eqnarray}
    \int_{0}^{t} \int_{0}^{t} C(t',t'') dt'dt'' &\le& \int_{0}^{t} \int_{0}^{t} \sigma ^{2} dt'dt'' \nonumber \\ &=& \sigma ^{2}t^{2},  
\end{eqnarray}
where $ \sigma $ represents the standard deviation of $ D(t) $ and is given by $ \sigma = \sqrt{\left\langle \left( D(t) - \mu \right) ^{2} \right\rangle} $. 
Thus, 
from Eqs.~(\ref{eq:SvarDcorr}) and (\ref{eq:betadvarS}), 
we have 
\begin{equation}
    A(t) \le \left( \frac{\sigma }{\mu } \right) ^{2}.  \label{eq:ulA}
\end{equation}
From this equation, 
we can see that when $ D(t) $ is a stationary process, 
$ A(t) $ is less than or equal to the square of the coefficient of variation of $ D(t) $. 
In addition, 
for a sufficiently small $ t $, 
we have 
\begin{equation}
    A(t) \approx \left( \frac{\sigma }{\mu } \right) ^{2} .  \label{eq:stA}
\end{equation}

When $ D(t) $ is ergodic, 
$ \lim_{t \to \infty } S(t) = \mu $. 
Thus, 
from Eq.~(\ref{eq:pdexp2gf}), 
when $ t \to \infty $, 
we have 
\begin{equation}
    G (\bm{x},t,\bm{x}_{0}) \sim \frac{1}{\left(4\pi \mu t \right)^{n/2}} \exp \left( -\frac{\left| \bm{x} -\bm{x}_{0} \right|^2}{4\mu t} \right). \label{eq:approxB}
\end{equation} 
In addition, 
from Eq.~(\ref{eq:betadvarS}), 
the non-Gaussian parameter $ A(t) $ converges to zero. 

Conversely,
if we have Eq.~(\ref{eq:approxB}) in the long time limit, 
$ D(t) $ is ergodic. 
When Eq.~(\ref{eq:approxB}) holds, 
from Eq.~(\ref{eq:pdexp2gf})
we have $ \lim_{t \to \infty } p(S,t) = \delta(S-\mu ) $. 
Thus, 
we have $ \lim_{t \to \infty } S(t)  = \mu $. 

In general, 
even if $ \lim_{t \to \infty } A(t) = 0 $, 
it does not necessarily mean that the propagator converges to a Gaussian distribution. 
However, 
for stationary stochastic DCs,  
the convergence of the non-Gaussian parameter to zero means that the propagator converges to a Gaussian distribution. 
From Eq.~(\ref{eq:betadvarS}), 
when $ \lim_{t \to \infty } A(t) = 0 $, 
we have $ \lim_{t \to \infty } S(t) = \mu $. 
When $ \lim_{t \to \infty } S(t) = \mu $, 
$ \lim_{t \to \infty } p(S,t) = \delta (S-\mu ) $. 
Thus, 
from Eq.~(\ref{eq:pdexp2gf}), 
we have Eq.~(\ref{eq:approxB}).

\subsection{Anomalous diffusion}    \label{ss:ad}
As shown in the previous section (Sec.~\ref{ss:nd}), 
when $ D(t) $ is a stationary process, 
the diffusion is normal. 
Thus, 
anomalous diffusion can occur only when $ D(t) $ is a nonstationary process. 
In particular, 
it is necessary that the process is a nonstationary process in which 
the ensemble average of the DC is time-dependent. 

Here, we assume that the ensemble average of the DC is given by 
\begin{equation}
    \left\langle D(t) \right\rangle = \alpha D_{c} t^{\alpha -1} \; (0 < \alpha < 1, 1 < \alpha ),    \label{eq:anmdass}
\end{equation}
where $ D_{c} $ is a positive constant. 
From Eq.~(\ref{eq:anmdass}), 
we have 
\begin{equation}
    \left\langle S(t) \right\rangle = D_{c} t^{\alpha -1}.  \label{eq:andS}
\end{equation}
Substituting this equation into Eq.~(\ref{eq:msdms}) leads to 
\begin{equation}
    \langle |\delta \bm{x}(t)|^{2} \rangle = 2nD_{c} t^{\alpha } .    \label{eq:defad}
\end{equation}
Thus, 
the diffusion is anomalous. 

Substituting Eq.~(\ref{eq:andS}) into Eq.~(\ref{eq:betavarS}) leads to 
\begin{eqnarray}
    A(t) &=& \frac{\left\langle \left( S\left( t \right) - D_{c}t^{\alpha -1} \right) ^{2}\right\rangle }{D_{c}^{2}t^{2(\alpha -1)}}    \\ 
          &=& \frac{\left\langle \left( S_{s} \left( t \right) - t^{\alpha -1} \right) ^{2}\right\rangle }{t^{2(\alpha -1)}},    \label{eq:andA}
\end{eqnarray} 
where $ S_{s} (t) = \frac{1}{t} \int _{0}^{t} D(t')/D_{c} dt' $.

\subsection{Case study}    \label{ss:cs}
In this section, 
we investigate properties of the propagators for two specific models of $ D(t) $. 
One is a simple model for which we can find an analytical expression for $ A(t) $, 
and the other is a realistic model. 
The simple model shows normal diffusion while the realistic model shows anomalous diffusion.  
For simplicity, 
the dimension is assumed to be one.

\subsubsection{Simple model}    \label{sss:sm}
Here, 
we assume that $ D(t) $ is described by a two-state Markov process. 
We also assume that the process is stationary. 
We label one state $ + $ and the other state $ - $. 
The DC is equal to $ D_{+} $ at $ + $ state and 
is equal to $ D_{-} $ at $ - $ state. 
The distribution of sojourn time in each state is given by 
\begin{eqnarray}
    \psi _{+} (\tau ) &=& \lambda _{-} e^{\lambda _{-}\tau },    \label{eq:edp} \\
    \psi _{-} (\tau ) &=& \lambda _{+} e^{\lambda _{+}\tau },    \label{eq:edm}
\end{eqnarray}
where $ \psi _{+} (\tau ) $ and $ \psi _{-} (\tau ) $ are 
the distributions of sojourn time in $ + $ state and $ - $ state, respectively. 
In Eqs.~(\ref{eq:edp}) and (\ref{eq:edm}), 
$ \lambda _{-} $ represents the transition probability from $ + $ state to $ - $ state and 
$ \lambda _{+} $ represents the transition probability from $ - $ state to $ + $ state. 
The initial distribution of the DC is the equilibrium distribution. 

\begin{figure}
\includegraphics[width=8.6cm]{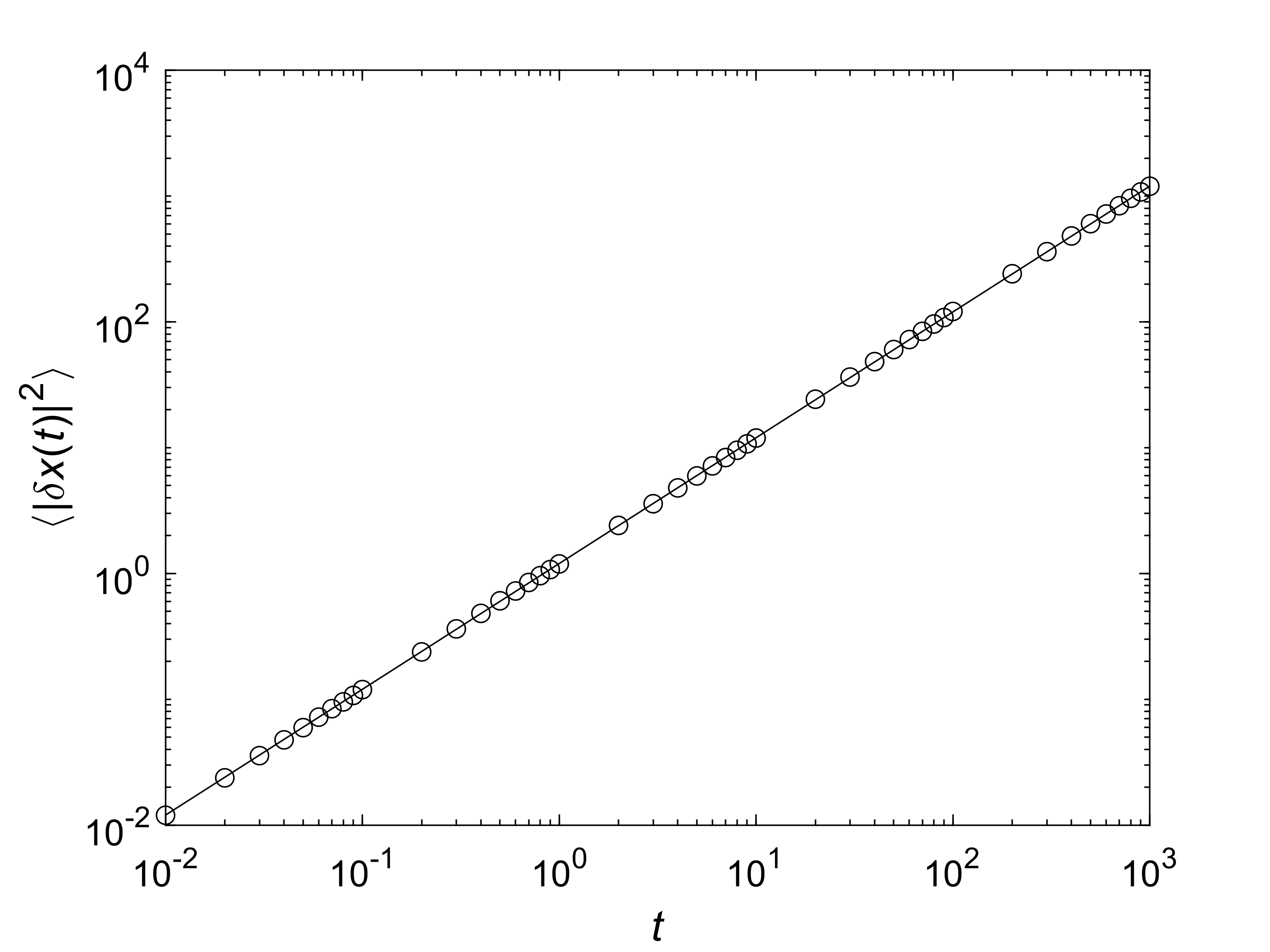}%
\caption{\label{fig:mM} Time dependence of the MSD. 
The solid line for the time dependence estimated from Eq.~(\ref{eq:thormsd});  
the open circles for the time dependence estimated from simulations. 
$ \lambda _{+} = 0.5 $, $ \lambda _{-} = 0.5 $, $ D_{+} = 1 $, $ D_{-} = 0.2 $. 
Time is normalized by $ t_{0} $ and the DC is normalized by $ D_{+} $. 
}
\end{figure}

For this model, 
we have 
\begin{eqnarray}
    \mu &=& \frac{D_{+} \lambda _{+} + D_{-} \lambda _{-}}{\lambda _{+} + \lambda _{-}},   \label{eq:mumarkov} \\
    \sigma ^{2} &=& \lambda _{+} \lambda _{-} \left( \frac{D_{+} - D_{-} }{\lambda _{+} + \lambda _{-}} \right) ^{2}. 
\end{eqnarray}
Thus, 
from Eq.~(\ref{eq:x2mu}), 
we have 
\begin{equation}
    \langle |\delta \bm{x}(t)|^{2} \rangle = 2 \frac{D_{+} \lambda _{+} + D_{-} \lambda _{-}}{\lambda _{+} + \lambda _{-}} t.  \label{eq:thormsd}
\end{equation}
Figure~\ref{fig:mM} demonstrates an excellent agreement between this result and the result estimated from simulations 
(see Appendix for the details of simulations). 

For the model, 
we can derive an analytical expression for $ A(t) $. 
For the model, 
we have the autocovariance function $ C(\Delta t) $: 
\begin{equation}
    C(\Delta t) = \sigma ^{2} e^{-\frac{\Delta t}{t_{0} }},    \label{eq:tmacov}
\end{equation}
where $ \Delta t $ is the time difference, and $ t_{0} $ is the time constant and is given by 
$ t_{0}  = 1/(\lambda _{+} + \lambda _{-}) $. 
Substituting Eq.~(\ref{eq:tmacov}) into Eq.~(\ref{eq:SvarDcorr}) leads to 
\begin{equation}
    \left\langle \left( S\left( t \right) - \mu \right) ^{2}\right\rangle = \frac{2\sigma ^{2} {t_{0}} ^{2}\left( \frac{t}{t_{0}}-e^{-\frac{t}{t_{0}}}-1 \right)}{t^{2}} .   \label{eq:SvarDcorrsp}
\end{equation}
By substituting this equation and Eq.~(\ref{eq:mumarkov}) into Eq.~(\ref{eq:betadvarS}), 
we have  
\begin{equation}
    A(t_{s}) = \frac{2C_{v}^{2}(t_{s}+e^{-t_{s}}-1)}{{t_{s}}^{2}},    \label{eq:Aexp}
\end{equation}
where $ t_{s} = t/t_{0} $, 
and $ C_{v} $ represents the coefficient of variation and is given by $ C_{v} = \sigma / \mu $. 
\begin{figure}
\includegraphics[width=8.6cm]{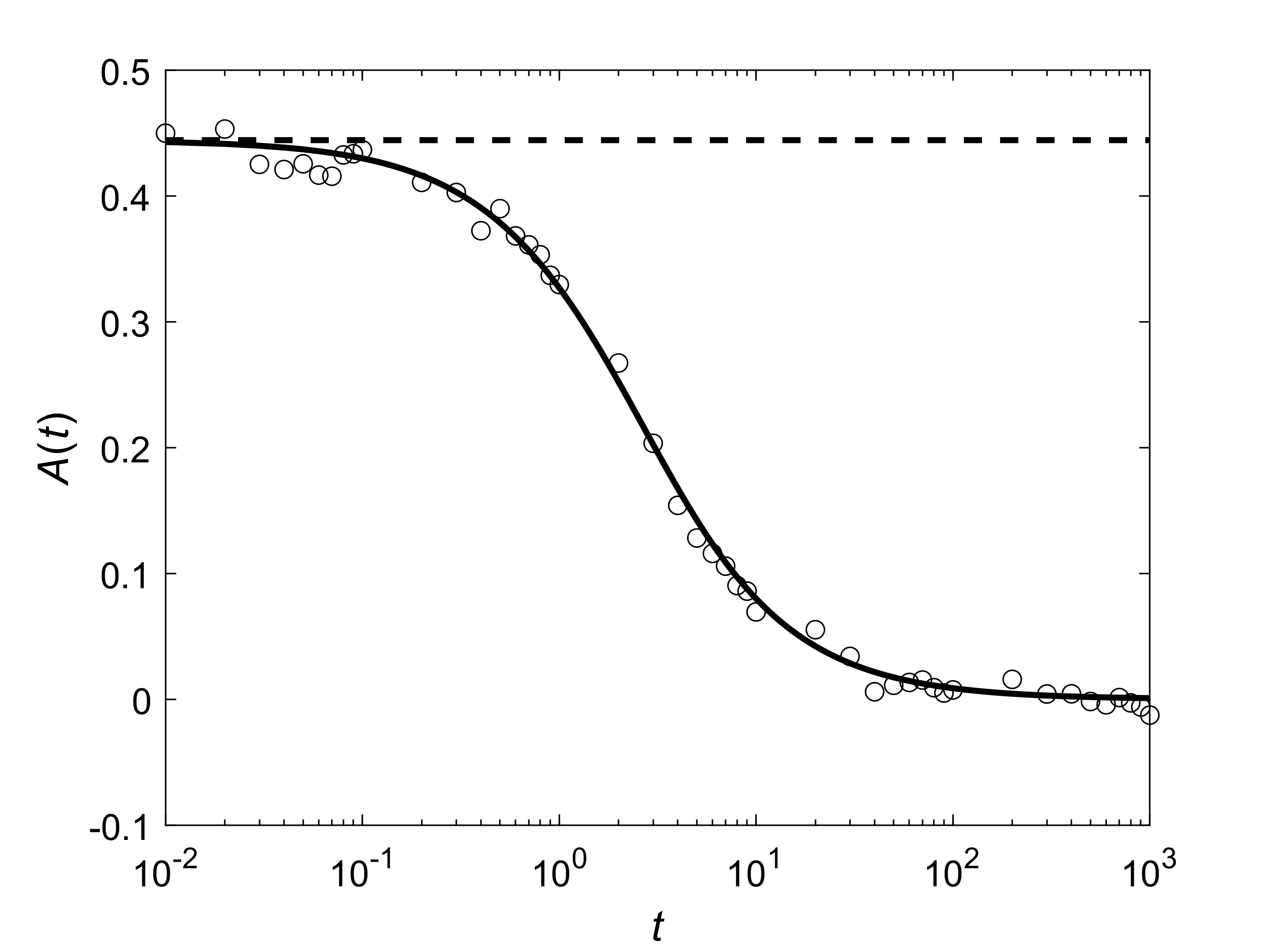}%
\caption{\label{fig:mA} Time dependence of the non-Gaussian parameter. 
The solid line for the time dependence estimated from Eq.~(\ref{eq:Aexp}); 
the open circles for the time dependence estimated from simulations; 
the dashed line for the upper limit estimated from Eq.~(\ref{eq:ulA}).   
}
\end{figure}
Figure~\ref{fig:mA} shows the time dependence of $ A(t) $. 
From this figure, 
we can see that 
the time dependence of the non-Gaussian parameter estimated from Eq.~(\ref{eq:Aexp}) is in good agreement with 
that estimated from simulations.  
We can also see that for short times, 
the value of the non-Gaussian parameter is nearly equal to $ C_{v}^{2} $, 
which is predicted from Eq.~(\ref{eq:stA}), but 
rapidly decreases to around zero on the time scale of one.   
This result indicates that on the time scale of one, 
a heavy-tailed propagator crosses over a distribution that is very close to a Gaussian distribution. 
\begin{figure}
\includegraphics[width=8.6cm]{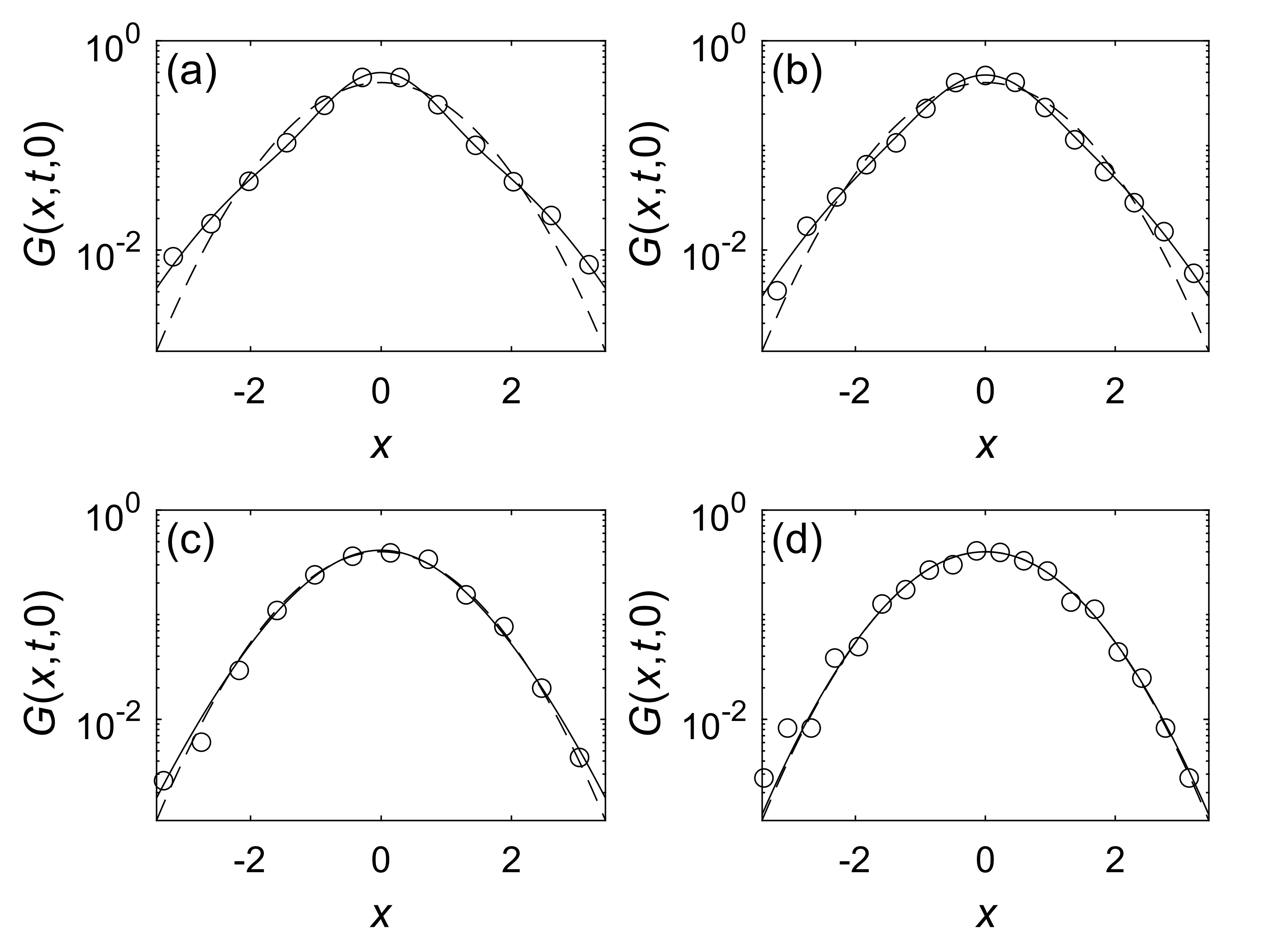}%
\caption{\label{fig:mG} Propagators at different times. 
The solid lines for the propagators estimated from Eq.~(\ref{eq:pdexp2gf}); 
the open circles for the propagators estimated from simulations; 
the dashed lines for the Gaussian distribution with the variance of one. 
The propagators are normalized so that their variances equal one.  
(a) $ t = 0.1 $. (b) $ t = 1 $. (c) $ t = 10 $. (d) $ t = 100 $. 
}
\end{figure}
Figure~\ref{fig:mG} shows that this is true. 
The time scale at which the crossover occurs is equal to the time scale of the correlation time of the DC.

\subsubsection{Realistic model}
Here, 
we examine the propagator of the model that describes subdiffusion of a molecule on living-cell membranes. 
The molecule is dendritic cell-specific intercellular adhesion molecule 3-grabbing nonintegrin (DC-SIGN), 
which is a receptor with unique pathogen-recognition capabilities.  
The model explains experimental data well~\cite{Manzo15}. 
The fluctuation of the DC comes from the dynamic heterogeneity of cell membrane. 

In the model, 
the distribution of the DC is given by a Gamma distribution:  
\begin{equation}
    P_{D}(D) = \frac{D^{\zeta -1}e^{-D/b}}{b^{\zeta }\Gamma (\zeta )}, 
\end{equation}
where $ P_{D}(D) $ represents the distribution of the DC, 
$ \Gamma (y) $ is the Gamma function, 
$ \zeta $ is the shape parameter, and 
$ b $ is the scale parameter.   
In addition, 
the conditional distribution of transit times $ \tau $ (the time when a molecule moves with a given $ D $) is given by an exponential distribution: 
\begin{equation}
    P_{\tau }(\tau | D) = \frac{D^{\gamma }}{l} e^{-\tau D^{\gamma }/l}, 
\end{equation}
where $ \gamma $ and $ l $ are constants. 
The initial distribution of the DC is $ P_{D}(D) $. 

\begin{figure}
\includegraphics[width=8.6cm]{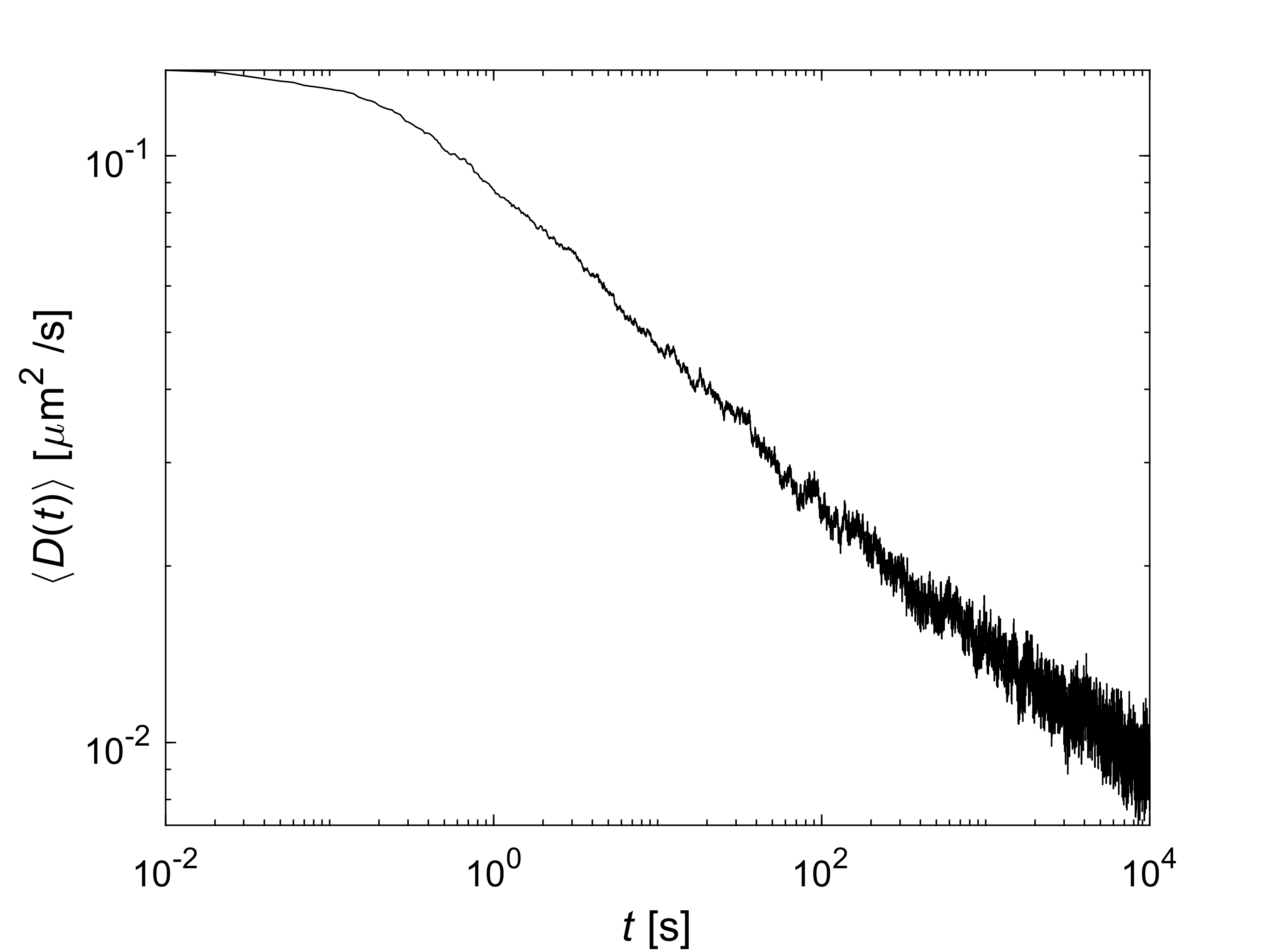}%
\caption{\label{fig:pD} Time dependence of the ensemble average of the DC. 
$ \zeta = 1.16 $, $ \gamma = 1.38 $, $ b = 0.12 $ $\mu $m$^{2} /$s, $ l = 0.10 $ $\mu $m$^{2\gamma } $s$^{\gamma +1} $. 
}
\end{figure}
In this model, 
the DC is a nonstationary process in which 
the ensemble average of the DC is time-dependent. 
Figure~\ref{fig:pD} shows the time dependence of $ \left\langle D(t) \right\rangle $. 
From this figure, 
we can see that the time dependence of $ \left\langle D(t) \right\rangle $ is given by Eq.~(\ref{eq:anmdass}) with $ 0 < \alpha < 1 $, 
although the value of $ \alpha $ changes around $ t = 1 $ s.  
This result indicates that diffusion is anomalous. 
\begin{figure}
\includegraphics[width=8.6cm]{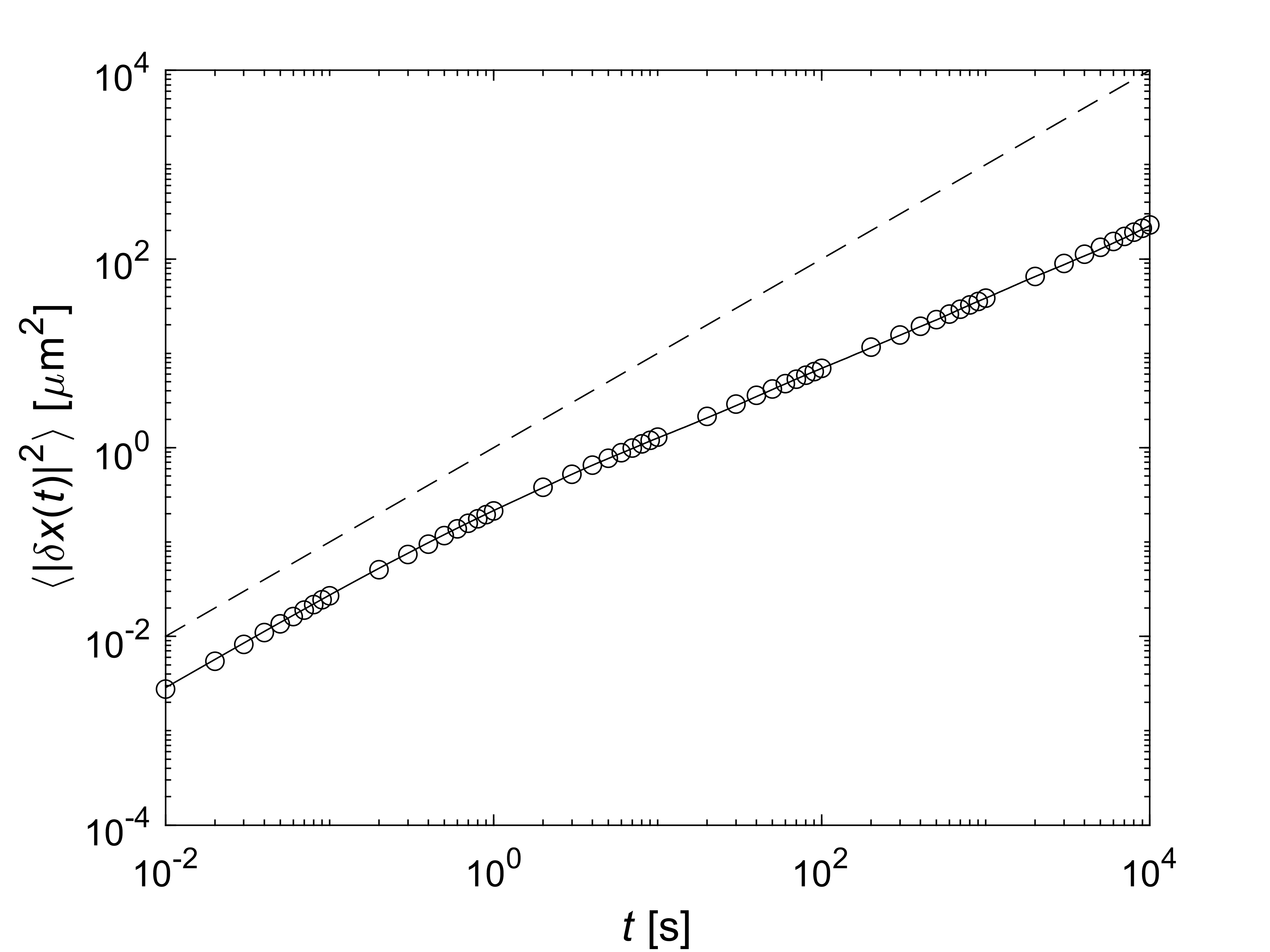}%
\caption{\label{fig:pM} Time dependence of the MSD. 
The solid line for the time dependence estimated from Eq.~(\ref{eq:msdms});  
the open circles for the time dependence estimated from simulations; 
the dashed line for the line with the slope of one.  
}
\end{figure}
Figure~\ref{fig:pM} shows the time dependence of $ \langle |\delta \bm{x}(t)|^{2} \rangle $. 
We can see that the time dependence of $ \langle |\delta \bm{x}(t)|^{2} \rangle $ is given by Eq.~(\ref{eq:defad}) with $ 0 < \alpha < 1 $, 
although the value of $ \alpha $ changes around $ t = 1 $ s.   
Thus, 
the diffusion is anomalous. 
From Fig.~\ref{fig:pM}, 
we can also see that the time dependence of $ \langle |\delta \bm{x}(t)|^{2} \rangle $ estimated from Eq.~(\ref{eq:msdms}) is in good agreement with 
that estimated from simulations. 

\begin{figure}
\includegraphics[width=8.6cm]{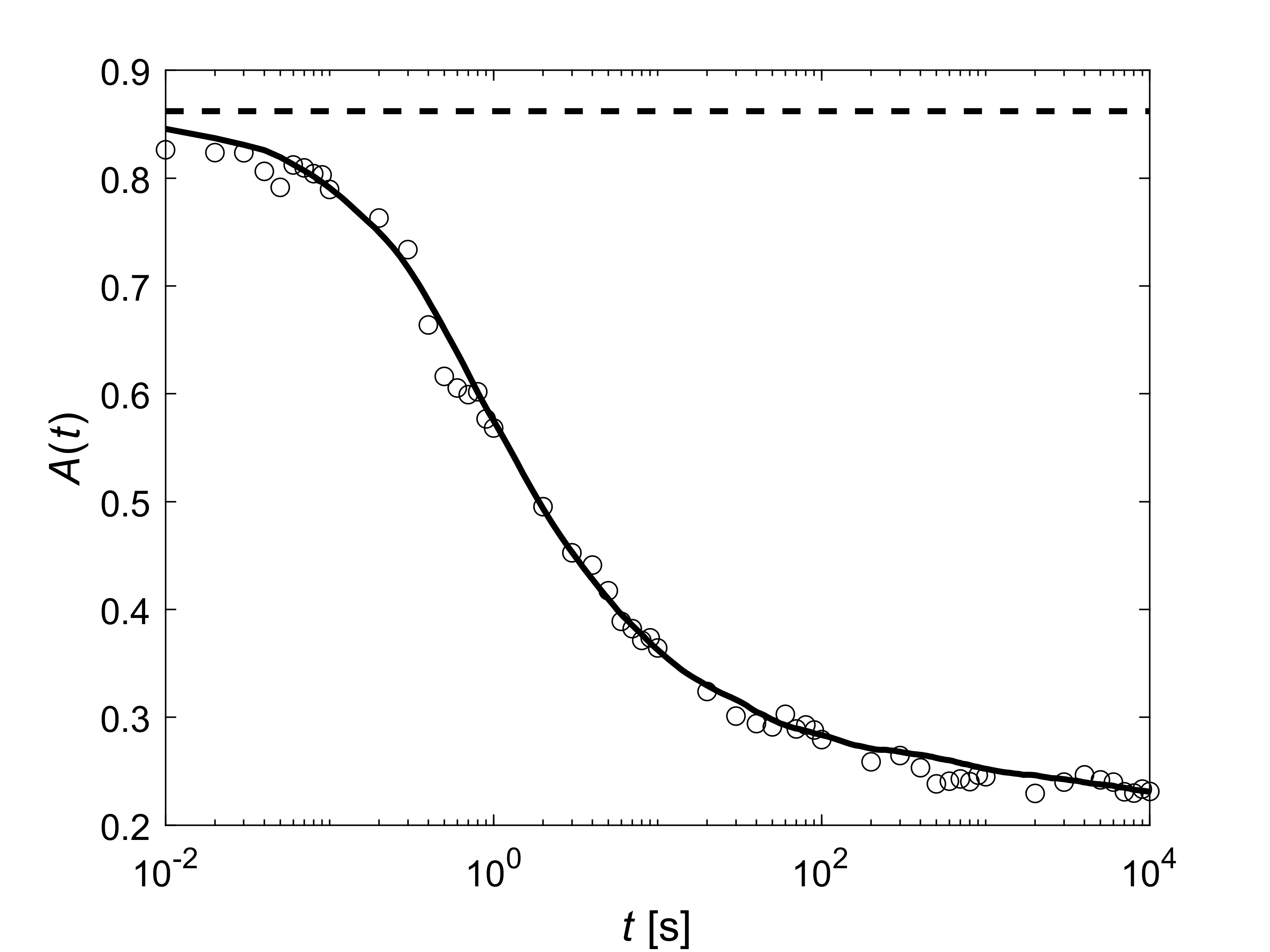}%
\caption{\label{fig:pA} Time dependence of the non-Gaussian parameter. 
The solid line for the time dependence estimated from Eq.~(\ref{eq:betavarS}); 
the open circles for the time dependence estimated from simulations; 
the dashed line for the value for short times estimated from Eq.~(\ref{eq:stAg}). 
}
\end{figure}
Figure~\ref{fig:pA} shows the time dependence of $ A(t) $. 
From this figure, 
we can see that 
the time dependence of the non-Gaussian parameter estimated from Eq.~(\ref{eq:betavarS}) is in good agreement with 
that estimated from simulations.  
In addition, 
for short times, 
the value of the non-Gaussian parameter is nearly equal to $ 1/\zeta $, 
which is predicted from Eq.~(\ref{eq:stAg}).  
From Fig.~\ref{fig:pA},  
we can also see that as in the two-state Markov model, 
the value of the non-Gaussian parameter 
rapidly decreases on the time scale of one.  
However, 
unlike the two-state Markov model, 
the non-Gaussian parameter continues to decrease slowly after reaching a certain positive value. 
This result indicates that the tails of the propagator become lighter on the time scale of one, 
but the tails still remain heavy.    
\begin{figure}
\includegraphics[width=8.6cm]{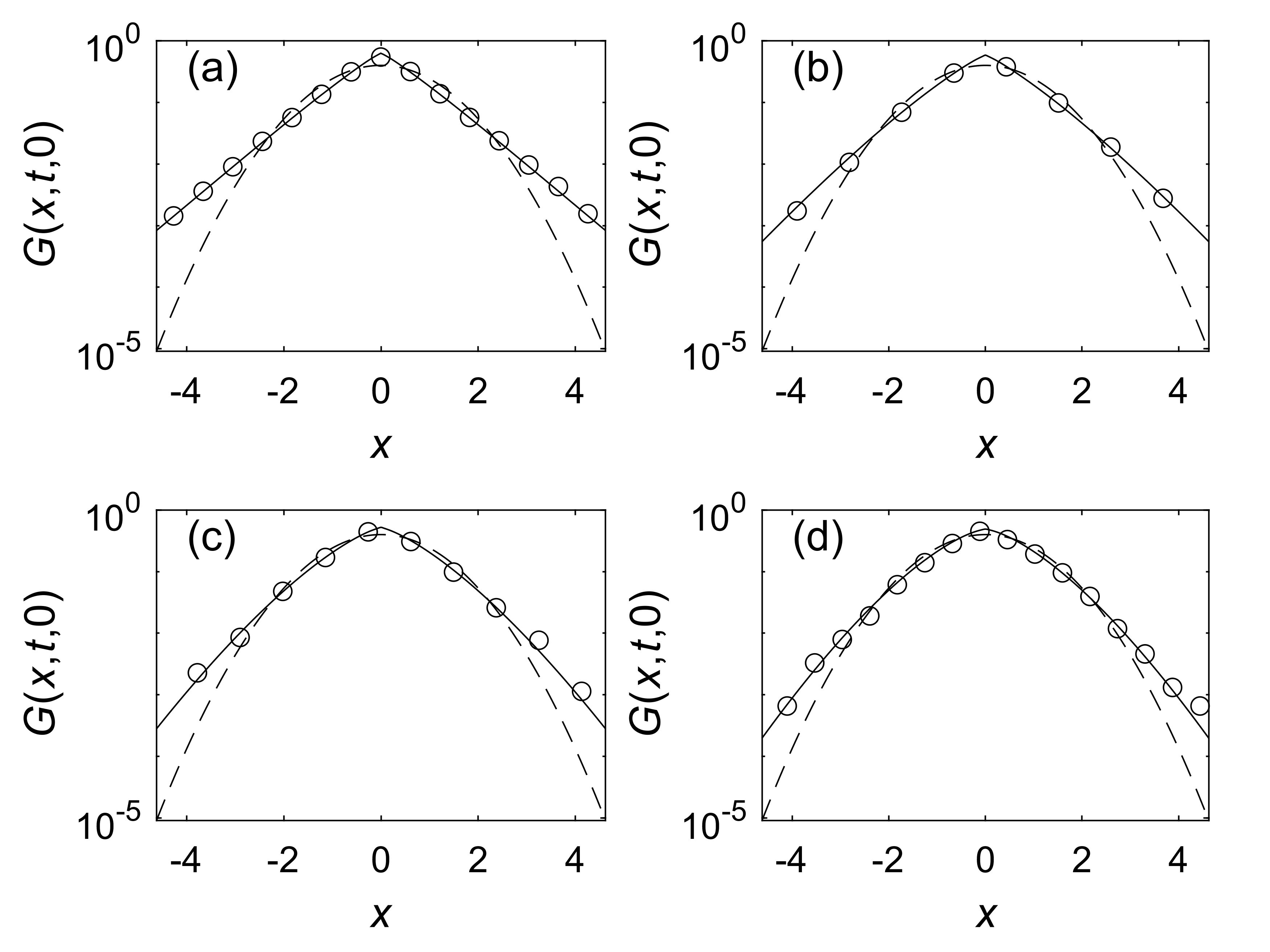}%
\caption{\label{fig:pG} Propagators at different times. 
The solid lines for the propagators estimated from Eq.~(\ref{eq:pdexp2gf}); 
the open circles for the propagators estimated from simulations; 
the dashed lines for the Gaussian distribution with the variance of one. 
The propagators are normalized so that their variances equal one.  
(a) $ t = 0.1 $ s. (b) $ t = 1 $ s. (c) $ t = 10 $ s. (d) $ t = 100 $ s. 
}
\end{figure}
Figure~\ref{fig:pG} shows that this is true.

\section{Discussion} \label{sec:discuss}
In the present study, 
we investigated properties of the diffusion with a broad class of stochastic DCs, 
focusing on the propagator.   
We showed that the propagator for the diffusing particle is unimodal and 
the peak is not sharp. 
We also showed that for a finite time,  
the propagator is non-Gaussian and the non-Gaussian parameter of the propagator is positive:  
the propagator is heavy-tailed relative to a Gaussian distribution. 
In addition, 
we showed that when a stochastic DC is ergodic, 
the propagator converges to a Gaussian distribution in the long time limit. 

An equation equivalent to Eq.~(\ref{eq:pdexp2gf}) can be derived using subordination~\cite{Chechkin17}. 
However, 
subordination has a different view than our approach. 
The essence of subordination is time change:  
subordination is the change of the time variable from real time to some kind of stochastic process. 
In subordination approach, 
the superensemble consists of ensembles with the same DC but different time flows. 
On the other hand, 
in our approach,  
the superensemble consists of ensembles with the same time flow but different time evolutions of TADC, 
which can be regarded as an effective DC. 

In a short time limit, 
Eq.~(\ref{eq:pdexp2gf}) has a relation with the propagator obtained by the superstatistical approach. 
In the superstatistical approach, 
the propagator $ G_{s} (\bm{x},t,\bm{x}_{0}) $ is given by 
\begin{equation}
    G_{s} (\bm{x},t,\bm{x}_{0}) = \int _{0}^{\infty }  \frac{r(D)}{\left(4\pi D t \right)^{n/2}} \exp \left( -\frac{\left| \bm{x} -\bm{x}_{0} \right|^2}{4D t } \right) dD , \label{eq:pdexp2gfss}
\end{equation} 
where $ r(D) $ is the probability distribution of $ D $~\cite{Beck03, Beck05, Beck06}. 
When $ t $ is sufficiently smaller than the time that characterizes the change in $ D(t) $, 
from Eq.~(\ref{eq:Mgt}), we have an approximation:
\begin{equation}
    S(\omega ) \approx D(0;\omega ) . \label{eq:Mgt3}
\end{equation} 
Thus, 
we have 
\begin{equation}
    G (\bm{x},t,\bm{x}_{0}) \approx \int _{0}^{\infty }  \frac{p(D, 0)}{\left(4\pi D t \right)^{n/2}} \exp \left( -\frac{\left| \bm{x} -\bm{x}_{0} \right|^2}{4D t } \right) dD . \label{eq:pdexp2gf3}
\end{equation} 
This is an approximation of the propagator by the superstatistical formula. 
For the minimal diffusing-diffusivity model, 
it has already been shown that 
when the time is sufficiently smaller than the correlation time of $ D(t) $, 
the propagator obtained from the subordination formula is equal to the propagator obtained by the superstatistical approach~\cite{Chechkin17}. 

Our approach is a natural extension of the superstatistical approach. 
This is because our approach includes the superstatistical approach as a special case. 
In Eq.~(\ref{eq:pifgt1}), 
if the DC is described by a special stochastic process in which each sample path of the DC is time independent, 
we can easily obtain Eq.~(\ref{eq:pdexp2gfss}) by our approach.  

In the present study, 
we showed that for a finite time, 
the propagator is heavy-tailed for stochastic DCs. 
The propagators of the anomalous diffusion described by CTRWs have a similar property: 
the propagators show exponential decay~\cite{Barkai20}.  
It has also been reported that anomalous diffusion described by CTRWs and anomalous diffusion with stochastic DCs have 
similar properties other than propagators~\cite{Miyaguchi16, Massignan14}. 
Thus, 
it is difficult to distinguish between anomalous diffusion described by CTRWs and anomalous diffusion with stochastic DCs. 
However, 
there is a case where there is a difference between the propagators of 
the diffusion with stochastic DCs and CTRWs. 
For one dimensional subdiffusion, 
the propagators of CTRWs often have a cusp at the origin and 
is not differentiable at the origin~\cite{Metzler00, Klafter11}. 
On the other hand, 
when $ x_{0} = 0 $ and 
$ t $ is larger than the time that characterizes the change of $ D(t) $, 
as we showed in Sec.~\ref{sec:nGGf}, 
the propagator in the Brownian motion with a stochastic DC has a smoother shape at the origin and 
is differentiable at the origin. 
In this sense, 
subdiffusion with stochastic DCs is different from that described by CTRWs. 

Convergence of the propagators to Gaussian distributions in the long time limit has been shown for 
specific stochastic models of DCs. 
For example, 
convergence of a heavy-tailed propagator to a Gaussian propagator has been shown for the minimal diffusing-diffusivity model 
with the equilibrium distribution of the DC as initial condition~\cite{Chechkin17,Sposini18}.  
Convergence to a Gaussian distribution has also been demonstrated  
for the two-state model with power-law distributions with the equilibrium distribution of the DC as initial condition~\cite{Miyaguchi16}. 
Our results suggest that the convergence in these models comes from the ergodicity of the models rather than 
characteristics unique to each model.   

In the present study, 
we showed that when DCs are ergodic, 
the propagators converge to Gaussian distributions in the long time limit. 
However, 
this does not necessarily mean that it takes an infinitely long time for the propagators to approach Gaussian distributions. 
From Eqs.~(\ref{eq:SvarDcorr}) and (\ref{eq:betadvarS}), 
we can see that it actually depends on how fast the autocovariance functions of DCs decay. 
In fact, 
when autocovariance functions decay exponentially, 
as in the simple model we used, 
$ A(t) $ approaches zero on the same time scale as the correlation time of the DC. 
On the other hand, 
if autocovariance functions decrease with power lows $ (\Delta t)^{-\kappa } (0 < \kappa < 1) $,
$ A(t) $ approaches zero at a rate of $ t^{-\kappa } $.  

In the present study, 
we showed that in a realistic model, 
the propagator does not converge to a Gaussian distribution even after 
10,000 s. 
However, 
it is unknown whether this has any physical meaning.  
This is because the consistency of the model with experimental data has been shown 
only up to a few seconds~\cite{Manzo15}, and it is unclear whether 
the model accurately describes diffusion of a molecule over longer times.  

In the present study, 
we showed that 
heavy-tailed propagators can be commonly observed in diffusion with stochastic DCs. 
This result may be important for triggered reactions in physics, chemistry, and biology. 
The result means that 
when a DC is described by a stationary process, 
some particles diffuse farther than in diffusion due to a Brownian motion with the constant DC that is equal to the mean of the fluctuating DC. 
In addition, 
if the MSDs are the same, 
some particles diffuse farther in anomalous diffusion with stochastic DCs 
than in anomalous diffusion due to a fractional Brownian motion, whose propagator is Gaussian~\cite{Deng09}.  
The property that some particles are transported farther in the same time is obviously important  
in diffusion-limited reactions triggered by single molecules.

\appendix*
\section{Simulations}    \label{app:sim}
We used the Euler method for numerical integration of the Langevin equation~\cite{Kloeden11}. 
\begin{equation}
    x(t+h) = x(t) + \sqrt{2D(t) h} \xi (t),  
\end{equation}
where $ h $ represents the time step in simulations. 
For the simple model, 
the time step in simulations is 0.001 up to time 1 and 0.1 after that. 
The number of trajectories we simulated is 40,000. 
For the realistic model, 
the time step in simulations is 0.001 s up to 1 s and 0.1 s after that.  
the number of trajectories we simulated is 40,000 for Figs.~\ref{fig:pM} and \ref{fig:pG} and 80,000 for Fig.~\ref{fig:pA}.

\begin{acknowledgments}
This work was supported by JSPS KAKENHI Grant Number 20H02058, 21K18679, 19H05718, and Tokyo Metropolitan Government Advanced Research Grant R2-2.
\end{acknowledgments}


\begin{thebibliography}{15}
\bibitem{Reingruber09} J. Reingruber and D. Holcman, Phys. Rev. Lett. {\bf 103}, 148102 (2009). 
\bibitem{Reingruber10} J. Reingruber and D. Holcman, J. Phys.: Condens. Matter {\bf 22}, 065103 (2010). 
\bibitem{Bressloff17} P. C. Bressloff and S. D. Lawley, J. Phys. A:Math. Theor. {\bf 50}, 195001 (2017). 
\bibitem{Manzo15} C. Manzo, J. A. Torreno-Pina, P. Massignan, G. J. Lapeyre, M. Lewenstein, and M. F. Garcia Parajo, Phys. Rev. X {\bf 5}, 011021 (2015). 
\bibitem{Chechkin17} A. V. Chechkin, F. Seno, R. Metzler, and I. M. Sokolov, Phys. Rev. X {\bf 7}, 021002 (2017). 
\bibitem{Metzler20} R. Metzler, Eur. Phys. J. {\bf 229}, 711 (2020). 
\bibitem{Miyaguchi16} T. Miyaguchi, T. Akimoto, and E. Yamamoto, Phys. Rev. E {\bf 94}, 012109 (2016). 
\bibitem{Barkai20} E. Barkai and S. Burov, Phys. Rev. Lett. {\bf 124}, 060603 (2020). 
\bibitem{Chubynsky14} M. V. Chubynsky and G. W. Slater, Phys. Rev. Lett. {\bf 113}, 098302 (2014). 
\bibitem{Molini11} A. Molini, P. Talkner, G. G. Katul, and A. Porporato, Physica A {\bf 390}, 1841 (2011). 
\bibitem{Soria21} M. Hidalgo-Soria, E. Barkai, and S. Burov, Entropy {\bf 23}, 231 (2021). 
\bibitem{Onuki98} R. Yamamoto and A. Onuki, Phys. Rev. Lett. {\bf 81}, 4915 (1998). 
\bibitem{Hofling13} F. H\"{o}fling and T. Franosch, Rep. Progr. Phys. {\bf 76}, 046602 (2013). 
\bibitem{Mardia70} K. V. Mardia, Biometrika {\bf 57}, 519 (1970). 
\bibitem{Beck03} C. Beck and E. G. D. Cohen, Physica A {\bf 332}, 267 (2003). 
\bibitem{Beck05} C. Beck, E. G. D. Cohen, and H. L. Swinney, Phys. Rev. E {\bf 72}, 056133 (2005). 
\bibitem{Beck06} C. Beck, Prog. Theoret. Phys. Suppl. {\bf 162}, 29 (2006). 
\bibitem{Massignan14} P. Massignan, C. Manzo, J. A. Torreno-Pina, M. F. Garcia-Parajo, M. Lewenstein, and G. J. Lapeyre, Phys. Rev. Lett. {\bf 112}, 150603 (2014). 
\bibitem{Klafter11} J. Klafter and I. M. Sokolov, {\it First Steps in Random Walks} (Oxford University Press, New York, NY, 2011). 
\bibitem{Metzler00} R. Metzler and J. Klafter, Phys. Rep. {\bf 339}, 1 (2000). 
\bibitem{Sposini18} V. Sposini, A. V. Chechkin, F. Seno, G. Pagnini, and R. Metzler, New J. Phys. {\bf 20}, 043044 (2018).  
\bibitem{Deng09} W. Deng and E. Barkai, Phys. Rev. E {\bf 79}, 011112 (2009). 
\bibitem{Kloeden11} P. E. Kloeden and E. Platen, {\it Numerical Solution of Stochastic Differential Equations} (Springer, Berlin, 2011). 
\end{thebibliography}

\end{document}